\documentclass[twocolumn,showpacs,superscriptaddress,10pt]{revtex4-1}
\usepackage{amsmath,amssymb,graphicx}
\usepackage{color}

\def\Ai{\qopname\relax{no}{Ai}}
\def\Ei{\qopname\relax{no}{Erfi}}

\begin{document}
\title{On the asymptotic evolution of finite energy Airy wavefunctions}

\author{Pedro Chamorro-Posada}
\email{pedcha@tel.uva.es}
\author{Julio S\'anchez-Curto}
\affiliation{Departamento de Teor\'{\i}a de la Se\~nal y Comunicaciones e Ingenier\'{\i}a Telem\'atica, Universidad de Valladolid, ETSI Telecomunicaci\'on, Paseo Bel\'en 15, 47011 Valladolid, Spain}
\author{Alejandro B. Aceves}
\affiliation{Department of Mathematics, Southern Methodist University, Clements Hall 221 ,  Dallas, TX  75275, USA}
\author{Graham S. McDonald}
\affiliation{Joule Physics Laboratory, School of Computing, Science and Engineering, Materials and Physics Research Centre, University of Salford, Salford M5 4WT, United Kingdom}

\begin{abstract}
In general, there is an inverse relation between the degree of localization of a wavefunction of a certain class and its transform representation dictated by the scaling property of the Fourier transform.  We report that in the case of finite energy Airy wavepackets a simultaneous increase in their localization in the direct and transform domains can be obtained as the apodization parameter is varied.  One consequence of this is that  the far field diffraction rate of a finite energy Airy beam decreases as the beam localization at the launch plane increases.  We analyse the asymptotic properties of finite energy Airy wavefunctions  using the stationary phase method.   We obtain one dominant contribution to the long term evolution that admits a Gaussian-like approximation, which displays the expected reduction of its broadening rate as the input localization is increased.

\end{abstract}

\maketitle

Airy wavefunctions \cite{berry} describe nondiffracting optical beams, nondispersing optical pulses in fibers or nonspreading quantum wavepackets. They have a number of remarkable properties.  As self-bending beams, their trajectories can be ballistically controled \cite{hu2010}.  Also, the main beam characteristics are recovered after the blockage of its central lobe along the propagation path.  This has been named the self-healing property of Airy beams \cite{healing}.   Among other applications, Airy beams have been proposed for particle micromanipulation \cite{micromanipulation}, the reduction of the beam scintillation due to atmospheric turbulence \cite{gu} or for nonlinear optical applications  \cite{chen2013, polynkin}.  The outstanding properties of Airy wavepackets are not limited to the field of optics and their interest extends to any physical system described by a Schr\"odinger-type evolution equation, such as in the realization of steerable electronic wavepackets \cite{Voloch2013}.

Pure Airy waveforms are associated with optical beams of infinite energy.  Some sort of truncation is required to produce finite energy solutions that keep the main characteristics of Airy beams over a limited propagation distance and can be used in experimental and theoretical investigations.  The exponential apodization of the initial beam profile proposed by Siviloglou and Christodoulides \cite{siviloglou} is the most widely used approximation to the infinite energy Airy beam.  

\begin{figure}[htbp]
\centering
\begin{tabular}{cc}
(a)&(b)\\
\includegraphics[width=.45\columnwidth]{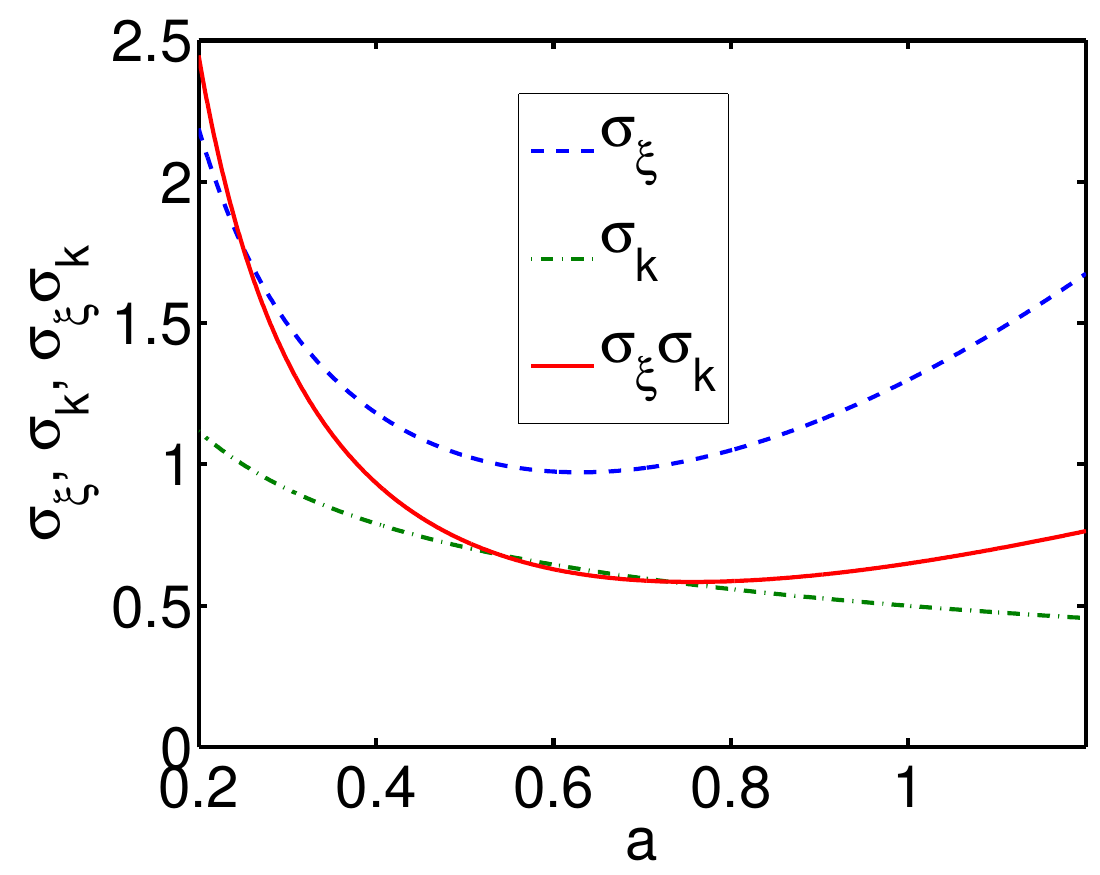}&
\includegraphics[width=.45\columnwidth]{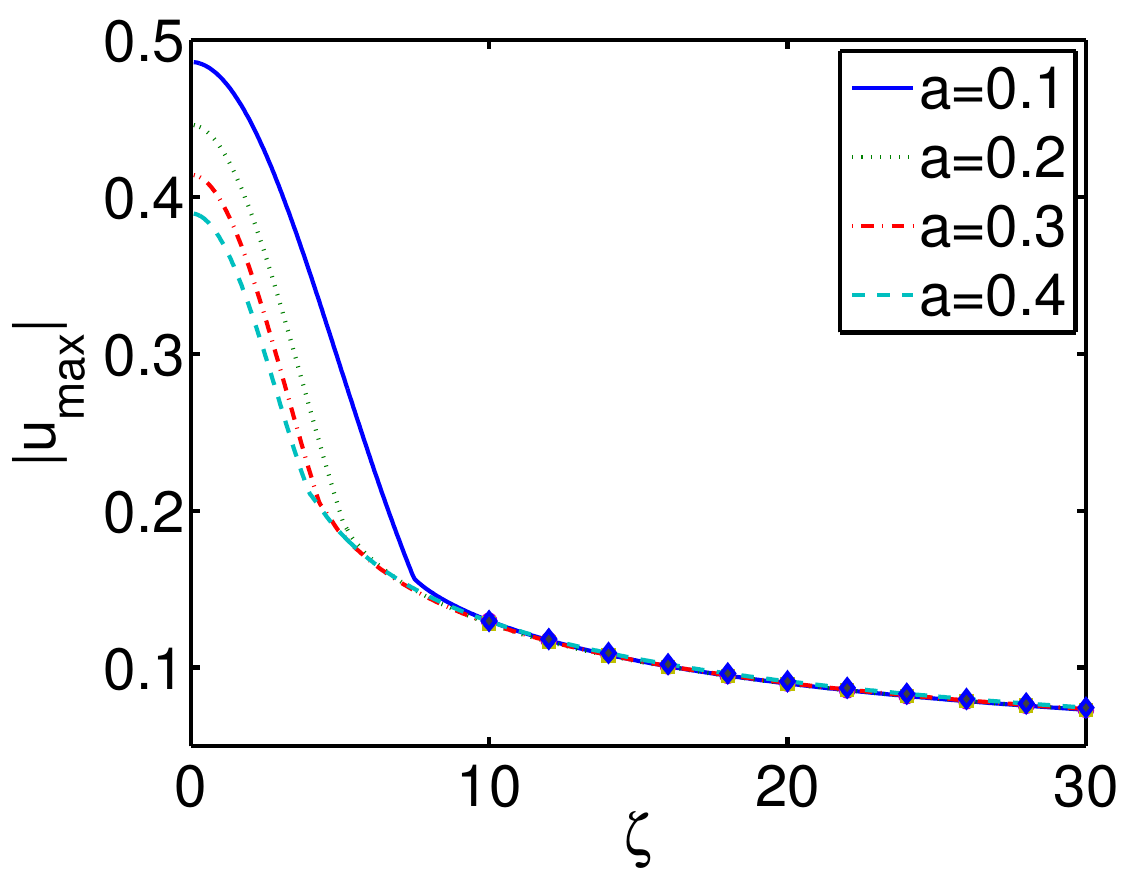}\\
(c)&\\
\includegraphics[width=.45\columnwidth]{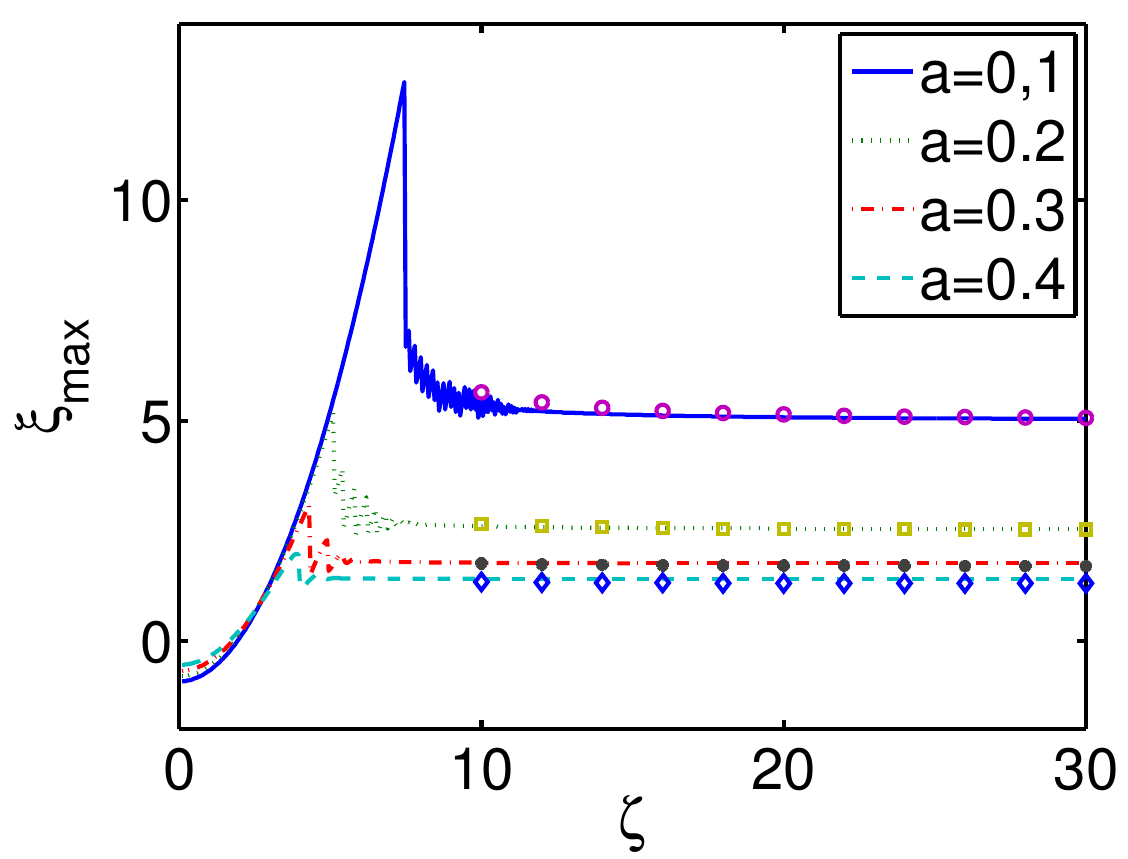}
\end{tabular}
\caption{ (a) RMS spatial $\sigma_\xi$  and spectral $\sigma_k$ widths of a finite-energy Airy function and $\sigma_\xi\sigma_k$  as a function of $a$.(b) Peak amplitude of the Airy beam (lines) and $u_a$ (points), (c) corresponding positions of beam maxima.}\label{maximo}
\end{figure}

In this letter, we highlight a new intriguing property of the finite energy Airy wavepackets.  Normally, the scaling property of the Fourier transform dictates that a higher localization of a wavepacket of a certain class results in a lower localization of its transform representation.  This is, for instance, the expected behavior of a Gaussian wavefunction as dictated by the uncertainty principle:  if we reduce the uncertainty of its position (time) the indetermination of its momentum (energy) increases.  For finite energy Airy wavefunctions this is not the case, and localization is increased simultaneously in its direct and transform representations,  always within the limits imposed by the uncertainty principle {shown in Figure \ref{maximo} (a) which we also proved analytically.}  For an optical beam, this has the associated remarkable consequence that the far field asymptotic diffraction broadening of the solution decreases as the spatial localization of the initial condition is increased.  

In this work, by use of asymptotic analysis we describe the evolution of finite energy Airy wavepackets.  While the results are discussed in terms of optical beam propagation, they equally apply to the dynamics of any system described by the Schr\"odinger equation.  Analyses of the propagation properties of Airy beams using asymptotic methods have been presented in \cite{kaganovsky,ring,ruipin,yiqing}.  Here we concentrate on characteristics of finite energy Airy beams obtained from the stationary phase method.  This approach permits to identify a particularly relevant contribution to the beam dynamics that admits a Gaussian-type approximation.  This approximate mode has Gaussian phase and amplitude, plus a nontrivial factor that produces a shift of the beam intensity peak position relative to that of the Gaussian term.  Then, as it was commented above, it is found that the localization in the far field is correlated with that of the initial condition whereas the asymptotic evolution of the beam intensity peak value is nearly the same at all degrees of initial localization.  We identify the regions of validity of this approximation that extend relatively close to the illumination plane. 

The Schr\"odinger equation $j\partial_\zeta u+1/2 \partial_{\xi\xi} u=0$  has an exact finite energy Airy beam solution\cite{siviloglou}
\begin{align}
\begin{split}
u(\xi,\zeta)&=\Ai\left(\xi-\left(\zeta/2\right)^2+ja\zeta\right)\\ 
&\times\exp\left(a\xi-a\zeta^2/2+j\left(-\zeta^3/12+a^2\zeta/2+\xi\zeta/2\right)\right)
\end{split} \label{eq:beam}
\end{align}
which reduces to a pure Airy beam \cite{berry} when $a=0$. For $a\ge 1$, the field at $\zeta=0$ and $\xi<0$ is negligible and the oscillations of the Airy function at negative values of the argument are effectively washed out from the initial condition, which becomes very close to a bell-shaped lobe with nearly constant phase. The rms width of the input beam achieves its minimum value at $a\simeq 0.63$.  We will be primarily interested in the behavior of these solutions at the smaller values of $a$ since the simultaneous localization increase in the direct and transform domains ceases to exist above a certain value of $a$, as discussed below.

\begin{figure}[htbp]
\centering
\begin{tabular}{cc}
(a)&(d)\\
\includegraphics[width=.4\columnwidth]{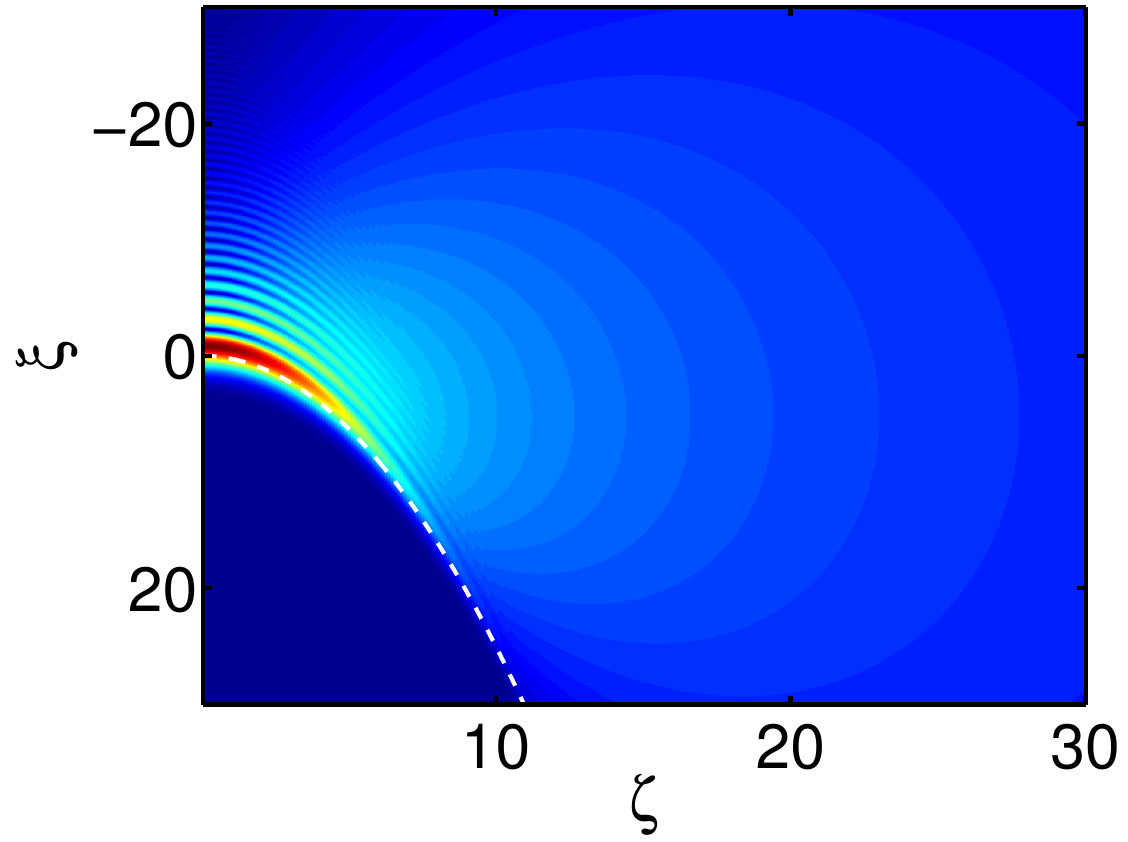}&\includegraphics[width=.4\columnwidth]{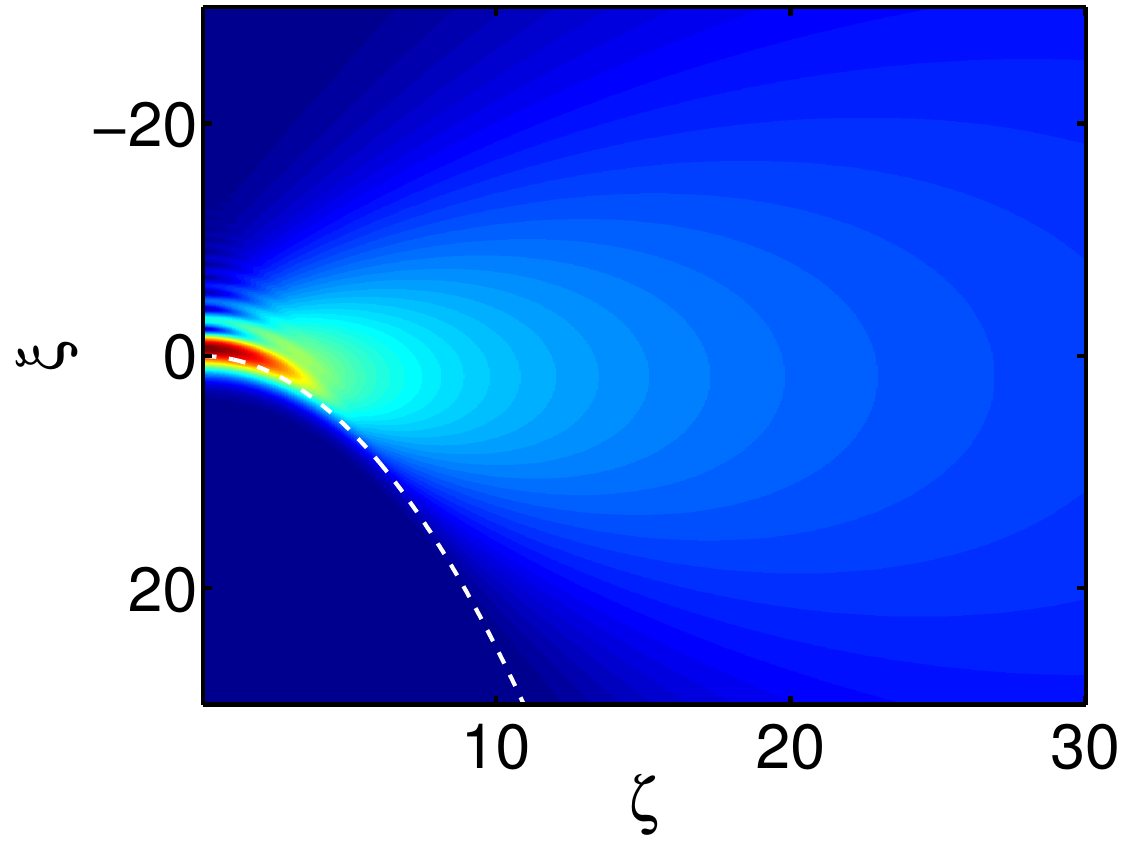}\\
(b)&(e)\\
\includegraphics[width=.4\columnwidth]{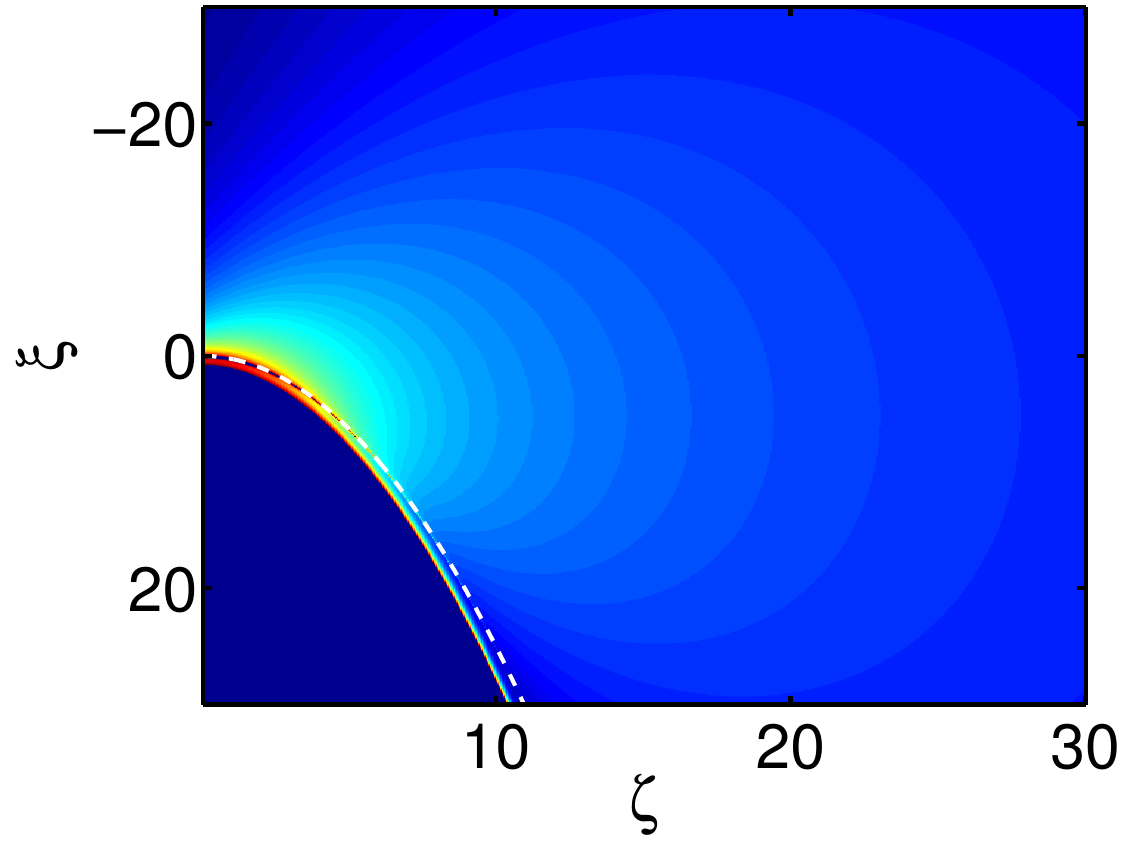}&\includegraphics[width=.4\columnwidth]{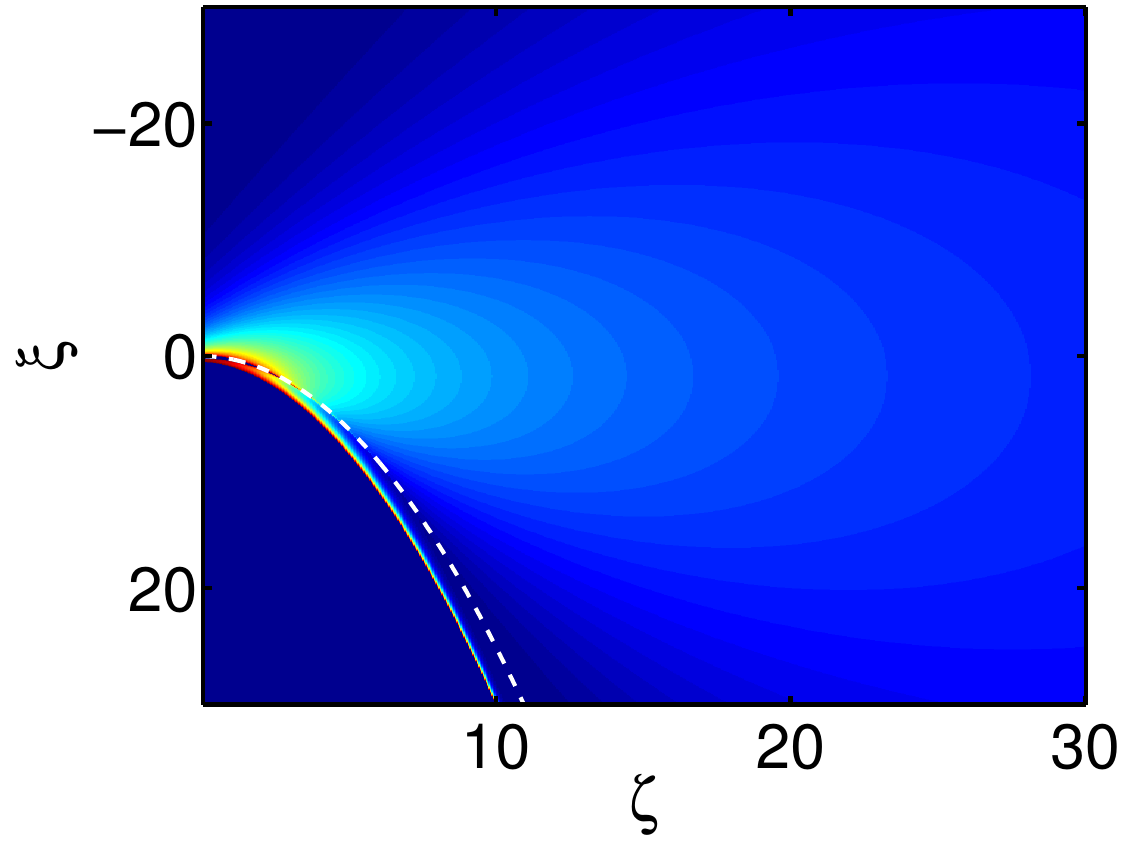}\\
(c)&(f)\\
\includegraphics[width=.4\columnwidth]{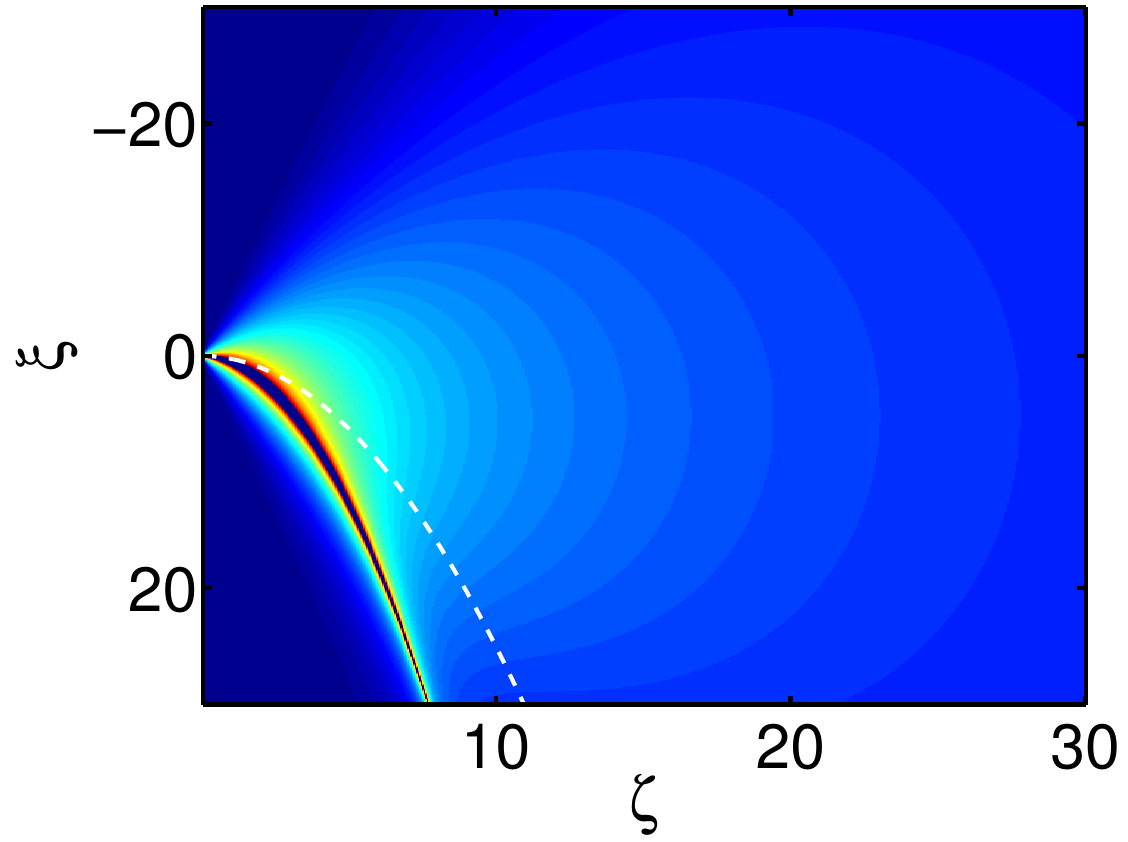}&\includegraphics[width=.4\columnwidth]{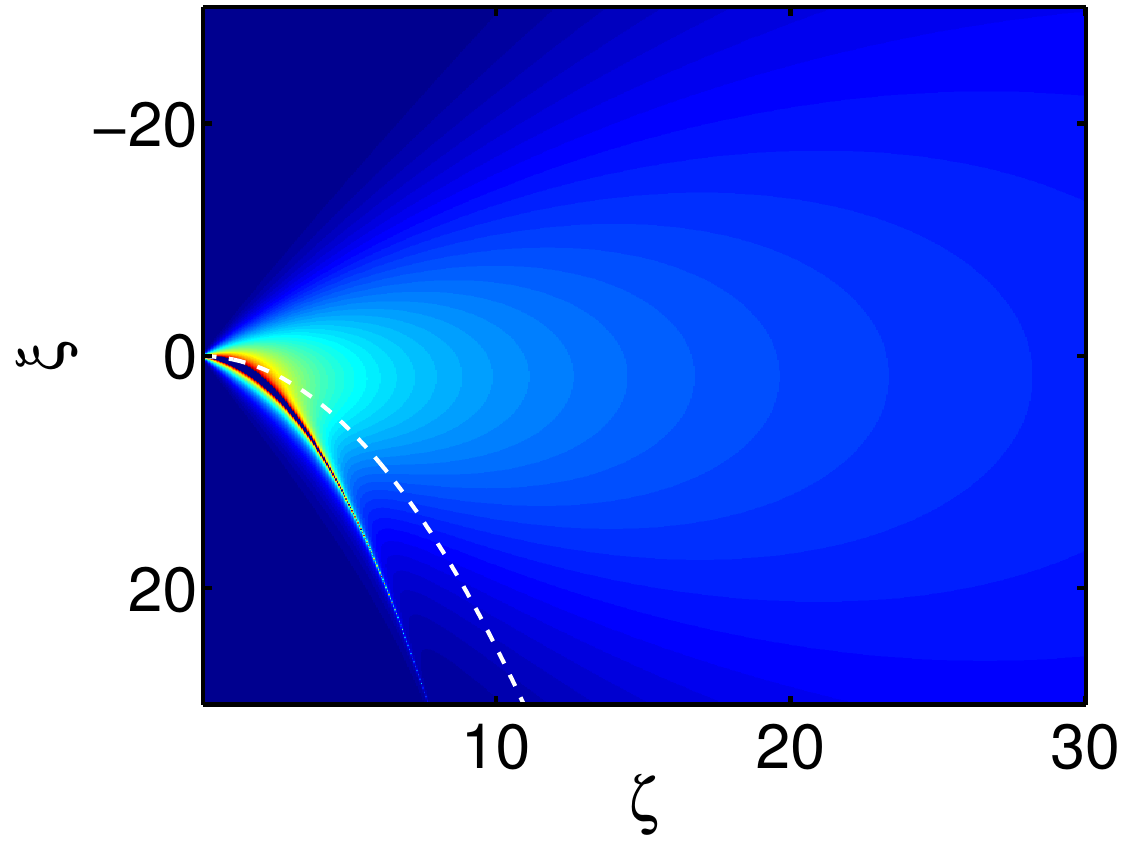}
\end{tabular}
\caption{(a) Amplitude of an exact Airy $a=0.1$ beam, (b) asymptotic term $u_1$  (c) the Gaussian-like mode. (d), (e) and (f) are the corresponding results for the $a=0.3$ beam. }\label{imagenesy1}
\end{figure}

In order to obtain the beam asymptotics we start with the initial condition $u(\xi,0)=\Ai\left(\xi\right)\exp\left(a\xi\right)$.  Its Fourier transform provides the angular spectrum 
\begin{equation}
 U(k_\xi,\zeta=0)=\exp\left(a^3/3-ak_\xi^2+j\dfrac{1}{3}\left(k_\xi^3-k_\xi a^2\right)\right).  \label{espectro}
\end{equation}
So, the degree of localization of the solution increases with $a$ simultaneously in the spatial and spectral domains. For an arbitrary $\zeta$, the beam Fourier transform is obtained by a direct multiplication with the Fresnel diffraction operator $U(k_\xi,\zeta)=U(k_\xi,\zeta=0)\exp\left(-jk_\xi^2\zeta/2\right)$ which, upon inverse transforming, gives the beam evolution
\begin{equation}
 u\left(\xi,\zeta\right)=\dfrac{\exp(a^3/3)}{2\pi}\int_{-\infty}^{\infty}\exp\left(-a k_\xi^2\right)\exp\left(j\zeta h\left(k_\xi;\xi,\zeta\right)\right)dk_\xi\label{eq:integral}
\end{equation}
with $h(k_\xi;\xi,\zeta)=-\dfrac{k_\xi^2}{2}+\dfrac{1}{\zeta}\left(k_\xi\xi+\dfrac{1}{3}\left(k_\xi^3-k_\xi a^2\right)\right)$.
For large $\zeta$, the integral in \eqref{eq:integral} can be evaluated asymptotically using the stationary phase method. We define $\delta=\left(\xi-a^2/3\right)/\zeta$.    

The two roots of $h'(k_\xi)=0$ 
\begin{equation}
c_l=\zeta/2\left(1+(-1)^l\sqrt{1-4\delta/\zeta}\right)\,\,\,\,l=1,2
\end{equation} 
give two contributions to the expansion $u(\xi,\zeta)\sim u_1(\xi,\zeta)+u_2(\xi,\zeta)$ at $\zeta\to\infty$ with 
\begin{equation}
 u_l=\dfrac{\exp\left(a^3/3+(-1)^{l}j\pi/4\right)}{\sqrt{2\pi}} \dfrac{\exp\left(-ac_l^2+j\zeta h(c_l)\right)}{\sqrt{(-1)^{l}\zeta h''(c_l)}},\label{asin}
\end{equation}
$h''(c_1)=-\sqrt{1-4\delta/\zeta}$ and $h''(c_2)=\sqrt{1-4\delta/\zeta}$,
where the prime has been used to denote the derivative of the function. { $h(c_l)=-c_l\left(\zeta/6+\delta/3-\delta/\zeta\right) +\delta\zeta/6$.}
Both $u_1$ and $u_2$ become singular at $\delta=\zeta/4$ where $h''(c_1)$ and $h''(c_2)$ are null.  This happens at the caustic  $\xi=a^2/3+\zeta^2/4$
of the Airy beam.  { For $a=0$, both stationary points produce complex conjugate contributions to the transverse standard Airy function at $\delta<\zeta/4$.  For $a>0$, the exponential $\exp(-ac_l^2)$ in \eqref{asin} creates an asymmetry in the the two terms.  For $a$ not too small (the asymptotic analysis will not hold if $a = O(1/\zeta)$ or smaller),} the contribution $u_2$ is highly localized at the caustic, it decays very fast as we move away from it and it is  negligible for most $(\xi,\zeta)$ at distances sufficiently far from the illumination plane. In this work we are primarily concerned with the behavior at long distances from the source in the region close to the optical axis, so we will use the results provided by the asymptotic approximation above the caustic,  where  $u(\xi,\zeta)\sim u_1(\xi,\zeta)$.

To the lowest order in $\delta$ $u_1$ can be approximated, {for $\zeta$ sufficiently large}, by a Gaussian-like distribution 
\begin{equation}
 u_a(\xi,\zeta)\simeq\dfrac{\exp\left(a^3/3-j\pi/4\right)}{\sqrt{2\pi \left(\zeta-2\delta\right)}} \exp\left(-a\delta^2\right)\exp\left(j\zeta \dfrac{\delta^2}{2}\right).\label{gauss}
\end{equation}

{ Figure \ref{imagenesy1} shows the amplitudes of the exact  finite energy Airy beams  \eqref{eq:beam} and the distributions of the dominant asymptotic term $u_1$ and the approximate Gaussian-like mode $u_a$  for $a=0.1$ and $a=0.3$. The caustic is marked as a dashed line in all plots and, in order to ease the visual comparison between the plots, the amplitude has been truncated to the maximum value reached in Figure \ref{imagenesy1} (a) in all cases.  Even though the asymptotic solutions in Figure \ref{imagenesy1}  are represented in the whole $(\xi,\zeta)$ plane they are strictly defined and meaningful only above the caustic that includes the far-field region of interest in the $\zeta$ direction.}

\begin{figure}[htbp]
\centering
\begin{tabular}{cc}
(a)&(b)\\
\includegraphics[width=.45\columnwidth]{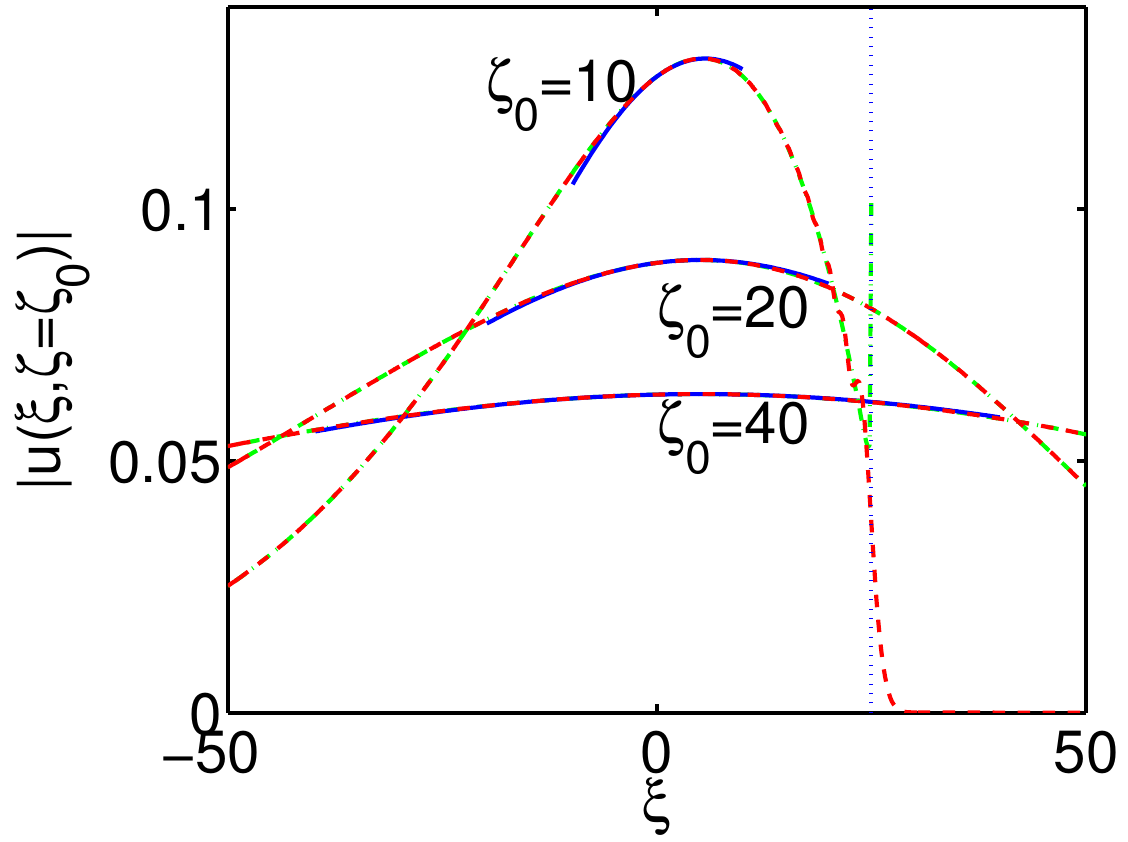}&
\includegraphics[width=.45\columnwidth]{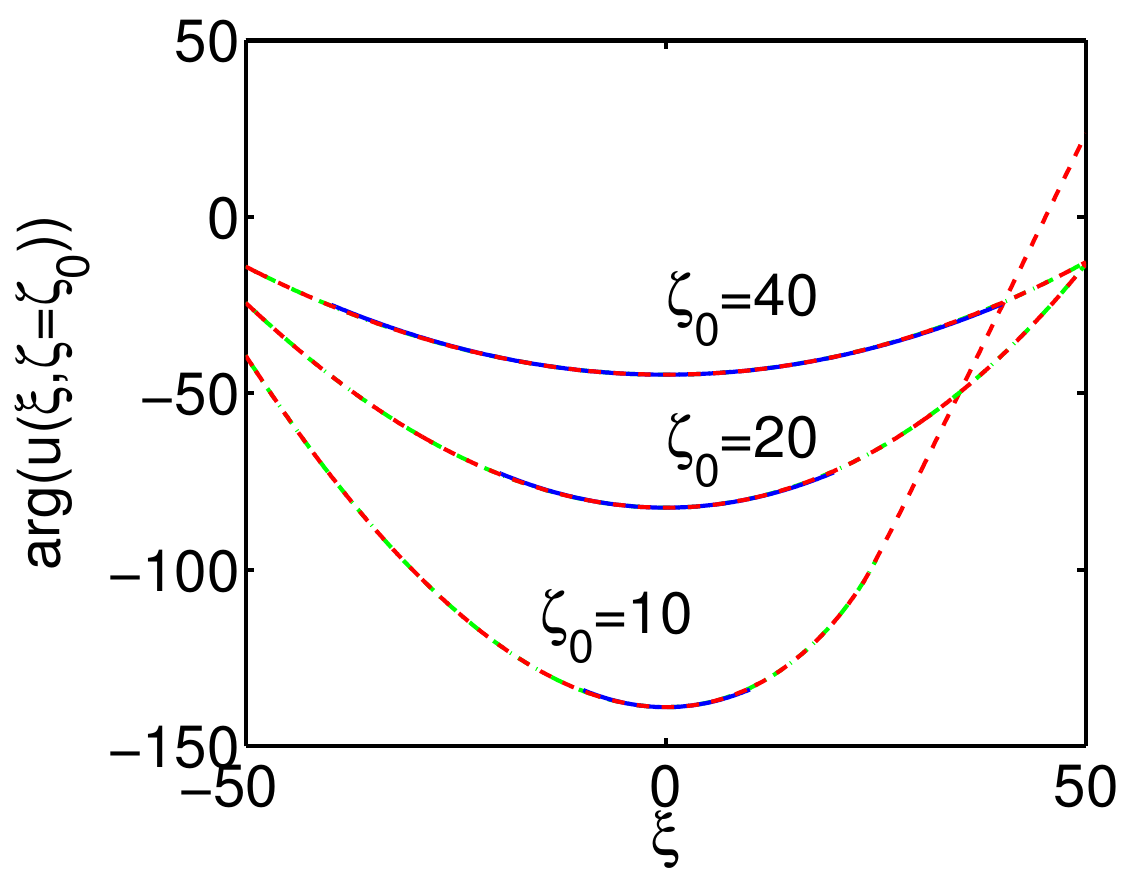}
\end{tabular}
\caption{ Amplitude (a) and phase (b), for $a=0.1$, of the exact Airy beam $u$ (dashed) $u_1$ (dashed-dotted green) and $u_a$ (solid) at three values of $\zeta_0$.  Vertical dotted line marks the caustic at $\zeta_0=10$. }\label{comparacion}
\end{figure}

\begin{figure}[htbp]
\centering
\begin{tabular}{cc}
(a)&(b)\\
\includegraphics[width=.45\columnwidth]{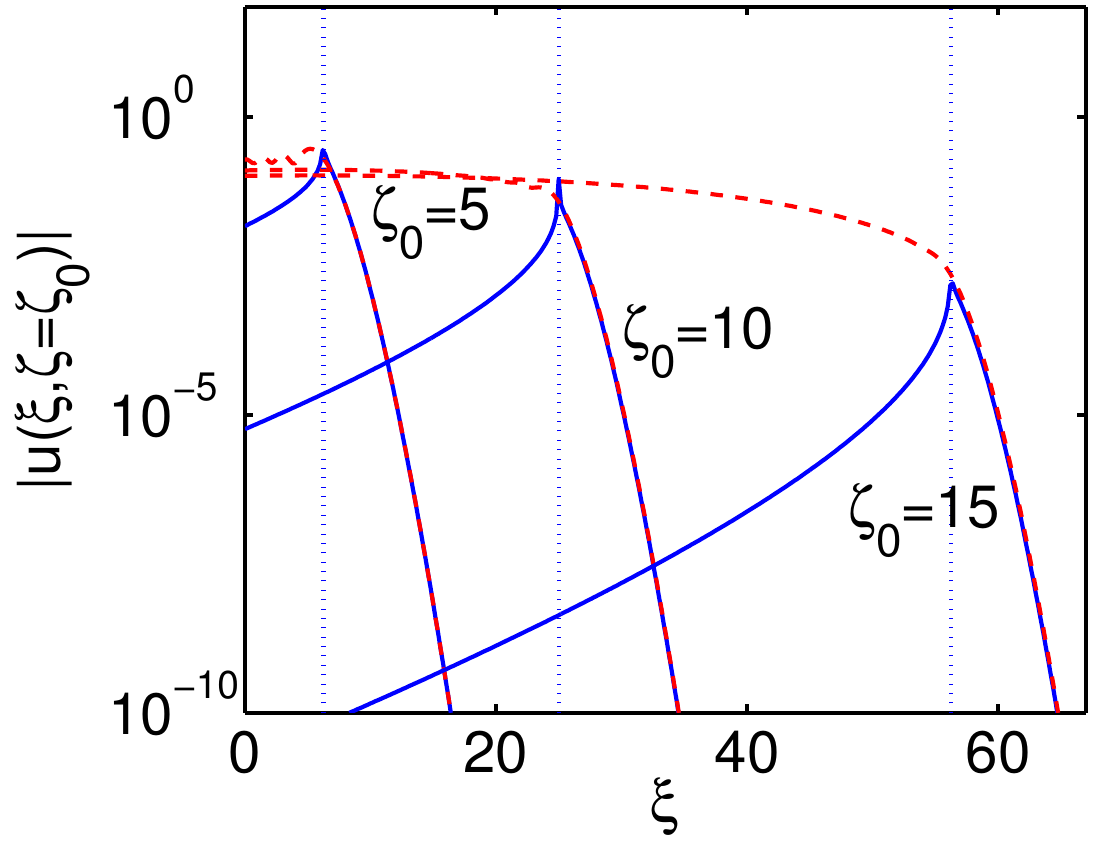}&
\includegraphics[width=.45\columnwidth]{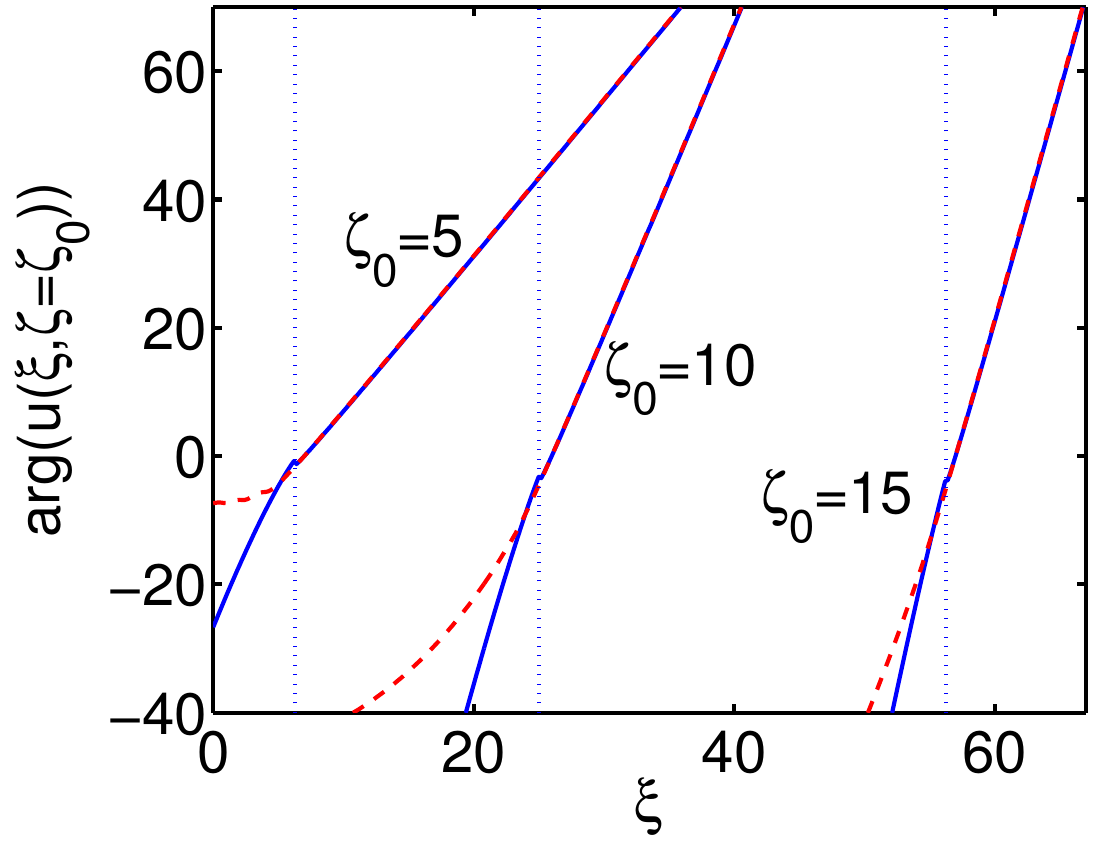}
\end{tabular}
\caption{ Amplitude (a) and phase (b), for $a=0.1$, of the exact Airy beam $u$ (dashed) and $u_2$ (solid)  at three values of  $\zeta_0$.  Vertical lines mark the caustic at each $\zeta_0$. }\label{comparacion2}
\end{figure}

The amplitudes and phases of the exact solution \eqref{eq:beam}, the asymptotic leading contribution above the caustic $u_1$ \eqref{asin} and the Gaussian-like mode \eqref{gauss} are compared in Figure \ref{comparacion} for three different propagation distances: $\zeta_0=10$, $20$ and $40$.  In all cases, the Gaussian-like approximation is plotted only for $|\delta|<1$ and, for $\zeta_0=10$, the asymptotic solution $u_1$ is displayed only above the beam caustic.  The figure shows that $u_1$ is an accurate representation of the beam at large $\zeta$ except when we are very close to the caustic and that the Gaussian-like approximation is very good for $|\delta|<1$.  

{ If $\delta>\zeta/4$ the stationary points are complex and the integral \eqref{eq:integral} must be evaluated using the steepest descent method \cite{felsen}.  Nevertheless, we note that the $u_2$ term correctly describes the beam asymptotics in this region as shown in Figure \ref{comparacion2} where the amplitudes (a) and phases (b) of the exact solution and $u_2$ \eqref{asin} are compared below the caustic.  The sign of the imaginary part of $h$ in $\exp(-j\zeta h(c_l))$ in \eqref{asin} at large $\zeta$ makes $u_1$ to diverge and $u_2$ to decrease exponentially in this region.}   The asymptotic behavior of the beam is then given by $u(\xi,\zeta)\sim u_1 (\xi,\zeta)$ if $\delta<\zeta/4$ and $u_2 (\xi,\zeta)$ if $\delta>\zeta/4$.  The asymptotic expansion is conventionally smoothed out close to the turning point with an Airy function \cite{felsen} that in our case is trivially the initial Airy beam.

Our study shows that the input beam width $\Delta \xi \simeq a^{-1}$ decreases as $a$ increases.  The spectral width of the input conditions is $\Delta k_{\xi} \simeq 1/\sqrt{2a}$ according to \eqref{espectro} and diminishes as the transverse localization of the beam increases.  Consistently, at a given $\zeta$  the beam width  of the Gaussian term that dominates the transverse localization properties of the beam is $\zeta/\sqrt{2a}$, showing the unexpected decrease of the diffraction induced beam broadening rate as the initial localization increases.  This is illustrated in Figure \ref{imagenesy1}.   Whereas $a=0.1$ imposes a lower degree of localization than $a=0.3$  in the input conditions, the far field beam broadening is slower in the $a=0.3$ case, with a stronger initial transverse confinement.  

The product $\Delta\xi\Delta k_{\xi}\simeq 1/\sqrt{2}a^{-3/2}$ decreases with $a$ until a minimum value is reached, always within the limits of the uncertainty principle. The minimum value of the product of the rms spatial and spectral widths is $\sigma_\xi\sigma_{k}\simeq0.83$ at $a\simeq 0.757$.  Therefore, the simultaneous localization increase in $\xi$ and $k$ spaces is limited to a given range of $a$.  For larger values of $a$, for which the oscillatory Airy tail is washed out, the rms spatial width starts to grow slightly faster than the inverse spectral rms width.  The existence of this limit permits to reconcile the space-momentum localization properties of finite-energy Airy wavefunctions with the uncertainty principle.  The dependence with $a$ of $\sigma_\xi$,$\sigma_{k_\xi}$ and their product is shown in Fig. \ref{maximo}(a).

Even though the transverse beam profile is dominated by the Gaussian term that peaks at $\xi=a^2/3$, the particular dependence of the square root in the denominator with $\zeta$ and $\xi$ shifts the beam center to $\xi_{max} \sim 1/(2a)+a^2/3$ for large $\zeta$ and $\delta<<\zeta$.  At the peak position, $\delta(\xi_{max},\zeta)=1/(2a\zeta)$.  Therefore, the asymptotic evolution of the peak amplitude is $\left|u_a(\xi=\xi_{max},\zeta)\right|\sim\exp\left(a^3/3\right)/\sqrt{2\pi\zeta}$. Since $\exp\left(a^3/3\right)\simeq 1$ for sufficiently small $a$, all the solutions will have very similar asymptotic peak amplitudes in spite of having different asymptotic beam widths;  the deviation is as small as $4.25\%$  for $a=0.5$ and $0.9\%$ for $a=0.3$.

Figure \ref{maximo} (b) shows the evolution of the finite energy Airy beam maxima for different values of $a$ with lines and the corresponding values for the approximate Gaussian mode with points.  The transverse positions corresponding to these maxima are shown in Fig. \ref{maximo}(c).  These plots display two clearly distinct regions in the beam evolution.  In the first stage, the beam peak follows the curved trajectory of the main lobe of the  Airy beam.  At a given point, both the trajectory described by the beam peak amplitude and its value start to proceed along the prediction of the quasi-Gaussian term.  We define this point $\zeta_f$ as an effective transition boundary that permits to quantify the domain of validity for the asymptotic approximation.  The dependence of $\zeta_f$ on the value of $a$ is very close to $1/a$ in the range of $a$ between $0.05$ and $0.4$.  For $\zeta>\zeta_f$ the beam is adequately described by the asymptotic term $u_1$ or its Gaussian approximation $u_a$. 

Although $u_a$ has a singularity in the transverse plane at each $\zeta$, its squared modulus can be integrated in the sense of the principal value of the integral and can be evaluated as the Hilbert transform of a Gaussian waveform to give
\begin{equation}
P(\zeta)=(\zeta/4)\exp\left(2a^3/3\right)\exp\left(-a\zeta^2/2\right)\Ei\left(\zeta\sqrt{a/2}\right)
\end{equation}
that, for large $\zeta$, reads $P\sim \exp\left(2a^3/3\right)/\sqrt{8\pi a}$
and coincides with the value of the power carried by the exact finite energy Airy beam \cite{siviloglou}.  { Function $\Ei$ is defined in terms of the error function as $\Ei(z)=-j\text{Erf}(j z)$.}

\begin{figure}[htbp]
\centering
\begin{tabular}{cc}
(a)&(b)\\
\includegraphics[width=.45\columnwidth]{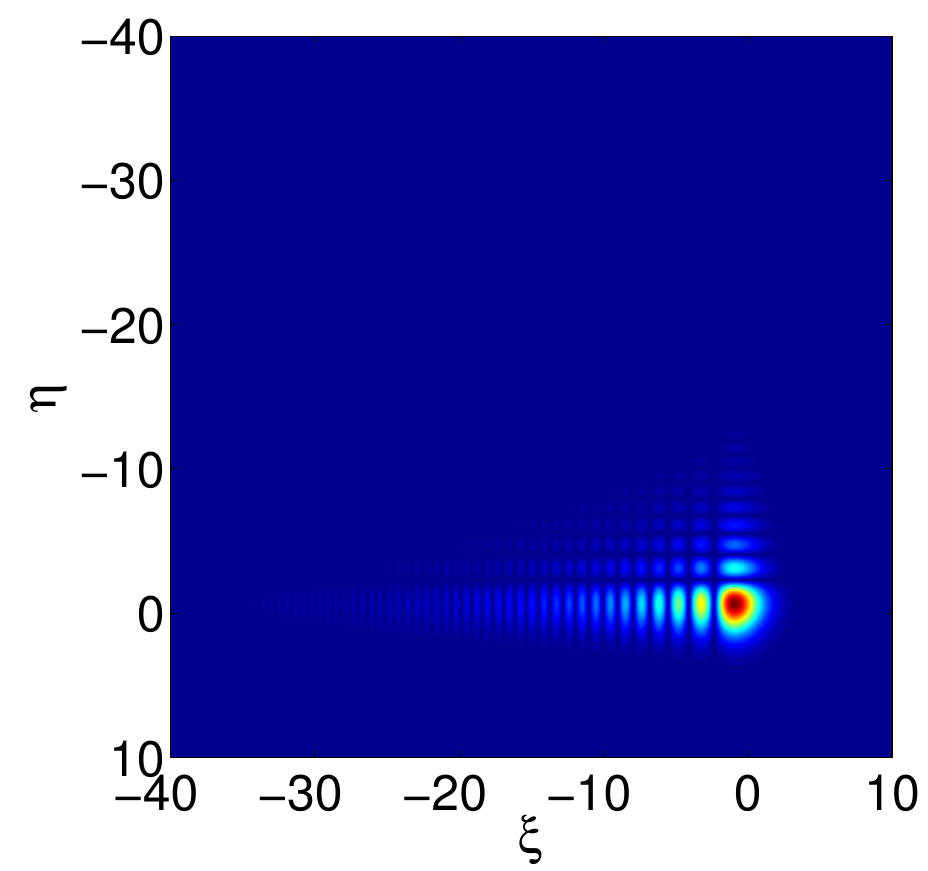}&
\includegraphics[width=.45\columnwidth]{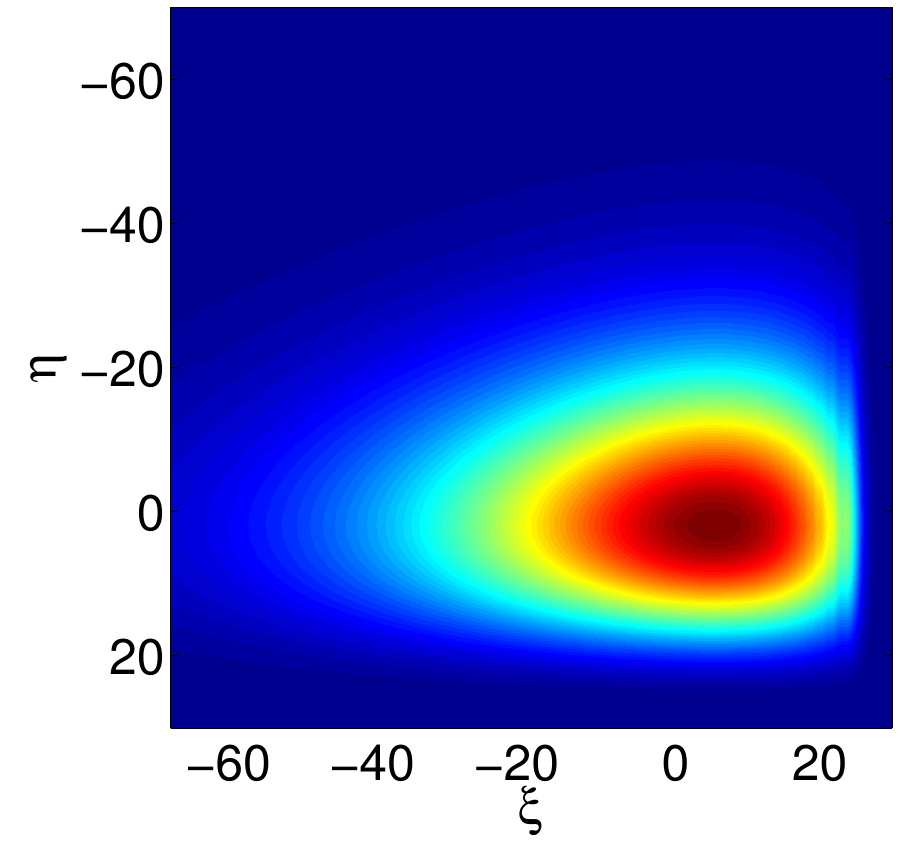}
\end{tabular}
\caption{Amplitude of a 2D Airy beam ($a=0.1$, $b=0.3$) at $\zeta=0$ and $\zeta=10$. }\label{2D}
\end{figure}

{Our results naturally apply to the 2D beam geometry when the beam is a product of two Airy functions, which is the solution of the 3D Fresnel equation $j\partial_\zeta w+1/2\partial_{\xi\xi} w+1/2 \partial_{\eta\eta} w=0$  given \cite{healing,yiqing} by $w(\xi,\eta,\zeta)=u(\xi,\zeta)v(\eta,\zeta)$ with $u(\xi,\zeta)$ in \eqref{eq:beam} and 
\begin{align}
\begin{split}
v(\eta,\zeta)&=\Ai\left(\eta-\left(\zeta/2\right)^2+jb\zeta\right)\exp\left(b\eta-b\zeta^2/2\right.\\
+&\left.j\left(-\zeta^3/12+b^2\zeta/2+\eta\zeta/2\right)\right),
\end{split} \label{eq:beamy}
\end{align}
where $b$ is the truncation parameter in the second $\eta$ transverse coordinate. The corresponding asympotic analysis follows inmediately by reproducing the results obtained for $u$ on the second term $v$. Figure \ref{2D} displays the input and far fied distributions of an transverse 2D beam with different values for $a$ and $b$ parameters.  Their axial ratios are shown not to display the relative inversion typically found in optical beam propagation.} 

In conclusion, we have presented an asymptotic analysis of finite energy Airy wavefunctions that provides an approximate analytical description of these solutions over extensive regions including those of interest in typical applications.  The results identify different evolution domains with markedly distinct behavior. {The enhancement of the nonlinear effects with the increase of the beam apodization reported in \cite{panos} could be linked to a manifestation of the \emph{linear} localization effects addressed in this work in the nonlinear regime}.   Our description can be very useful in problems such as the analysis of Airy beams at nonlinear interfaces \cite{chamorro}.  An effective study can be based on a plane-wave decomposition in the linear case \cite{chremmos}.  { The asymptotic analysis provides an useful description of the main component of the input beam in the interaction with a linear-nonlinear interface.}  The extension of the analysis to 2D highlights the existence of a correlation in the degree of localization in the initial and asymptotic evolution regions of finite energy Airy wavefunctions. {The observed behavior has necessarily to be linked with the complexity of the input beam and the interplay between the structure of the beam intensity and phase defining its wavefront.  This relation could be further explored, for instance, using intensity transport equation methods \cite{woods}.}

\end{document}